\input harvmac
\input epsf
\input amssym
%
\noblackbox
\newcount\figno
\figno=0
\def\fig#1#2#3{
\par\begingroup\parindent=0pt\leftskip=1cm\rightskip=1cm\parindent=0pt
\baselineskip=11pt
\global\advance\figno by 1
\midinsert
\epsfxsize=#3
\centerline{\epsfbox{#2}}
\vskip -21pt
{\bf Fig.\ \the\figno: } #1\par
\endinsert\endgroup\par
}
\def\figlabel#1{\xdef#1{\the\figno}}
\def\encadremath#1{\vbox{\hrule\hbox{\vrule\kern8pt\vbox{\kern8pt
\hbox{$\displaystyle #1$}\kern8pt}
\kern8pt\vrule}\hrule}}

\def\frac#1#2{{#1 \over #2}}

\def\semi{\subset\kern-1em\times\;}

\def\sqr#1#2{{\vcenter{\vbox{\hrule height.#2pt
\hbox{\vrule width.#2pt height#1pt \kern#1pt \vrule width.#2pt}
\hrule height.#2pt}}}}

     \def\cO{{\cal O}}
%

%

%

%

%

%

%
\def\coeff#1#2{\relax{\textstyle {#1 \over #2}}\displaystyle}
\def\ZZ{\Bbb{Z}}

\def\IR{\Bbb{R}}


%
\lref\frascati{ I.~Bena, Lectures given at  ``Winter School on Attractor Mechanism,'' INFN-Laboratori Nazionali di Frascati, Italy, March 2006, to appear.}
%
\lref\BenaDE{
I.~Bena and N.~P.~Warner,
Adv. Theor. Math. Phys. 9 (2006) 1-35
[arXiv:hep-th/0408106].
}
%
\lref\BenaTD{
  I.~Bena, C.~W.~Wang and N.~P.~Warner,
  JHEP {\bf 0603}, 015 (2006)
  [arXiv:hep-th/0411072].
}
\lref\BenaVA{
  I.~Bena and N.~P.~Warner,
  arXiv:hep-th/0505166.
}
%
\lref\BerglundVB{
  P.~Berglund, E.~G.~Gimon and T.~S.~Levi,
  arXiv:hep-th/0505167.
}
%
\lref\BenaZY{
  I.~Bena, C.~W.~Wang and N.~P.~Warner,
  arXiv:hep-th/0512157.
 }
%
\lref\GiustoIP{
  S.~Giusto, S.~D.~Mathur and A.~Saxena,
  Nucl.\ Phys.\ B {\bf 710}, 425 (2005)
  [arXiv:hep-th/0406103].
  S.~Giusto and S.~D.~Mathur,
  Nucl.\ Phys.\ B {\bf 729}, 203 (2005)
  [arXiv:hep-th/0409067].
}
%
\lref\BreckenridgeIS{
J.~C.~Breckenridge, R.~C.~Myers, A.~W.~Peet and C.~Vafa,
Phys.\ Lett.\ B {\bf 391}, 93 (1997)
[arXiv:hep-th/9602065].
}
%
\lref\GauntlettNW{
J.~P.~Gauntlett, J.~B.~Gutowski, C.~M.~Hull, S.~Pakis and
H.~S.~Reall,
Class.\ Quant.\ Grav.\  {\bf 20}, 4587 (2003)
[arXiv:hep-th/0209114].
%
}
\lref\GutowskiYV{
J.~B.~Gutowski and H.~S.~Reall,
JHEP {\bf 0404}, 048 (2004)
[arXiv:hep-th/0401129].
}
%
\lref\BenaTK{
  I.~Bena and P.~Kraus,
  JHEP {\bf 0412}, 070 (2004)
  [arXiv:hep-th/0408186].
}
\lref\BenaWT{
I.~Bena and P.~Kraus,
Phys.\ Rev.\ D {\bf 70}, 046003 (2004)
[arXiv:hep-th/0402144].
}
\lref\GauntlettQY{
  J.~P.~Gauntlett and J.~B.~Gutowski,
  Phys.\ Rev.\ D {\bf 71}, 045002 (2005)
  [arXiv:hep-th/0408122].
}

\lref\ElvangDS{H.~Elvang, R.~Emparan, D.~Mateos and H.~S.~Reall,
  Phys.\ Rev.\ D {\bf 71}, 024033 (2005)
  [arXiv:hep-th/0408120].
}
\lref\ElvangRT{H.~Elvang, R.~Emparan, D.~Mateos and H.~S.~Reall,
  Phys.\ Rev.\ Lett.\  {\bf 93}, 211302 (2004)
  [arXiv:hep-th/0407065].
}
\lref\newBWW{
  I.~Bena, C.~W.~Wang and N.~P.~Warner,
  arXiv:hep-th/0608217.
}
\lref\PalmerGU{
  B.~C.~Palmer and D.~Marolf,
  JHEP {\bf 0406}, 028 (2004)
  [arXiv:hep-th/0403025].
}
\lref\GreenSP{
  M.~B.~Green, J.~H.~Schwarz and E.~Witten,
``Superstring Theory. Vol. 1: Introduction,''
}
%
\lref\DenefRU{
  F.~Denef,
  JHEP {\bf 0210}, 023 (2002)
  [arXiv:hep-th/0206072].
}
%
\lref\DenefNB{
  F.~Denef,
  JHEP {\bf 0008}, 050 (2000)
  [arXiv:hep-th/0005049].
}
%
\lref\BatesVX{
  B.~Bates and F.~Denef,
  arXiv:hep-th/0304094.
}
%
\lref\KalloshVY{
  R.~Kallosh, A.~Rajaraman and W.~K.~Wong,
  Phys.\ Rev.\ D {\bf 55}, 3246 (1997)
  [arXiv:hep-th/9611094].
}
\lref\MathurZP{
  S.~D.~Mathur,
  Fortsch.\ Phys.\  {\bf 53}, 793 (2005)
  [arXiv:hep-th/0502050].
}
%
\lref\GrantQC{
  L.~Grant, L.~Maoz, J.~Marsano, K.~Papadodimas and V.~S.~Rychkov,
  JHEP {\bf 0508}, 025 (2005)
  [arXiv:hep-th/0505079].
}
%
\lref\BenaAY{
  I.~Bena and P.~Kraus,
  Phys.\ Rev.\ D {\bf 72}, 025007 (2005)
  [arXiv:hep-th/0503053].
}
\lref\GiustoZI{
  S.~Giusto, S.~D.~Mathur and Y.~K.~Srivastava,
  arXiv:hep-th/0601193.
}
%
\lref\SaxenaUK{
  A.~Saxena, G.~Potvin, S.~Giusto and A.~W.~Peet,
  arXiv:hep-th/0509214.
}
%
\lref\RychkovJI{
  V.~S.~Rychkov,
  JHEP {\bf 0601}, 063 (2006)
  [arXiv:hep-th/0512053].
}
%
\lref\Bak{
  D.~Bak, Y.~Hyakutake and N.~Ohta,
  Nucl.\ Phys.\ B {\bf 696}, 251 (2004)
  [arXiv:hep-th/0404104].
 D.~Bak, Y.~Hyakutake, S.~Kim and N.~Ohta,
  Nucl.\ Phys.\ B {\bf 712}, 115 (2005)
  [arXiv:hep-th/0407253].
}



\Title{
\vbox{
\hbox{\baselineskip12pt \vbox{\hbox{hep-th/0604110}
\hbox{NSF-KITP-06-25}
\hbox{CERN-PH-TH/2006-045 }}
}}}
{\vbox{\vskip -1.5cm
\centerline{\hbox{The Foaming Three-Charge Black Hole}}}}
\vskip -.3cm
\centerline{Iosif~Bena${}^{(1)}$,  Chih-Wei Wang${}^{(2)}$ and
Nicholas P.\ Warner${}^{(2,3)}$}

\bigskip
\centerline{{${}^{(1)}$\it School of Natural Sciences,
Institute for Advanced Study }}
\centerline{{\it Einstein Dr., Princeton, NJ 08540, USA }}
\medskip
\centerline{{${}^{(2)}$\it Department of Physics and Astronomy,
University of Southern California}}
\centerline{{\it Los Angeles, CA 90089-0484, USA}}
\medskip
\centerline{{${}^{(3)}$\it Department of Physics, Theory Division}}
\centerline{{\it CERN, Geneva, Switzerland}}
\medskip

\bigskip
\bigskip

We find a very large set of smooth horizonless geometries that have
the same charges and angular momenta as the five-dimensional,
maximally-spinning, three-charge, BPS black hole ($J^2 = Q^3 $).  Our
solutions are constructed using a four-dimensional Gibbons-Hawking
base space that has a very large number of two-cycles. The entropy of
our solutions is proportional to $\sqrt{Q}$. In the same class of
solutions we also find microstates corresponding to zero-entropy black
rings, and these are related to the microstates of the black hole by
continuous deformations.

\vskip .3in
\Date{\sl {April, 2006}}


\newsec{Introduction}

It has been expected for some time that there are many
supergravity solutions that have the same charges and angular
momenta as a given three-charge black hole or black ring. However,
the true extent of this degeneracy has only become apparent as the
result of recent work \refs{ \BenaVA, \BerglundVB}.  In these
papers it was shown that, through a geometric transition called a
``bubbling transition,'' one can generate a huge number of smooth,
horizonless supergravity solutions with a fixed set of asymptotic
charges and angular momenta. All these solutions are dual to
states in the D1-D5 conformal field theory (CFT), that describes
black holes and black rings. The existence of such a large number
of smooth horizonless solutions raises the distinct possibility
that the typical CFT states that contribute to the black-hole
entropy are dual to such solutions.  If this is true, it would
imply that the black hole should be thought of as an ``ensemble''
of horizonless solutions; this would greatly deepen our
understanding of black holes, and quantum gravity in general. A
review of this can be found in \refs{\MathurZP,\frascati}

In this letter we focus on the ``bubbling''  of
maximally-rotating, three-charge BPS black holes and black rings
in five dimensions.  We find a very large number of smooth
solutions that have exactly the same supersymmetries, size,
charges and angular momenta as these objects. Our solutions have
no localized brane sources (all the charges come from fluxes
wrapping non-trivial cycles), and no horizons, and thus can be
thought of as (a subset of) the ``microstates'' of the
corresponding classical objects.

The equations underlying our solutions were found in
\refs{\GauntlettNW,\GutowskiYV,\BenaDE,\GauntlettQY}.  The
``bubbling transition'' is obtained by replacing the flat $\IR^4$,
which constitutes the four-dimensional spatial base of the
five-dimensional black-hole and black-ring solutions, by a half-flat
four-dimensional
(hyper-K\"ahler) metric whose signature is allowed to change between
$+4$ and $-4$.  For computational convenience, we take the base to be
a four-dimensional Gibbons-Hawking (GH) space with many centers and
whose  geometric charges are integers of any sign.  To be asymptotic
to $\IR^4$, the sum of the geometric charges must be equal to one, but
near a negative charge the metric of the base will become
negative-definite.  As shown in \refs{\GiustoIP,\BenaVA,\BerglundVB},
even if the four-dimensional base changes its overall sign, the warp
factors ensure that full eleven-dimensional solution is smooth and
Lorentzian.  The GH base metrics have non-trivial two-cycles that run
between the geometric charges; these are the ``bubbles,'' and the
bubbling transition may be thought of as a process of pair creation,
separation and decomposition of geometric charge.  In this way the
charges of the branes are ``topologically partitioned'' amongst the
bubbles.

The three-charge, rotating, BPS  black hole in five dimensions
(also known as the the BMPV black hole) has two equal angular
momenta that are bounded above by the square root of the product
of the charges: $J_1=J_2 \leq \sqrt{Q_1 Q_5 Q_P}$.   The
maximally-spinning (or zero-entropy) BPS black hole saturates this
bound.  We show that if we have a large enough number of bubbles
(or cycles),  with fluxes of comparable magnitude, one always
obtains $J_1^2=J_2^2 \approx Q_1 Q_5 Q_P$ in the limit when the
number of bubbles becomes large. In fact, if the number of centers
is large enough, it appears much more difficult to find a solution
that does not have $J_1^2=J_2^2 \approx Q_1 Q_5 Q_P$. To violate
this,  some of the fluxes have to be very much larger than
others.

If the fluxes on all the cycles that end on one of the
positive-charge GH points are very much larger than all the other
fluxes, then that GH point will move very far  from the rest of
the GH points, and the solutions become one of the microstates of
zero-entropy BPS black rings
\refs{\BenaWT,\ElvangRT,\BenaDE,\ElvangDS,\GauntlettQY}\foot{The
solutions that we consider here are more general than the
zero-entropy black ring microstates constructed in \BenaVA: The
latter solutions involved a single geometric transition that
resulted in two bubbles, while here we consider clusters of large
numbers of bubbles. Other black ring microstate geometries have
been constructed in \GiustoZI.}. In fact, there is no sharp
distinction between black-ring microstates and black-hole
microstates: When the positive-charge GH point is very far away
from the others then the solution is clearly a ring microstate and
when this point is in a cluster with all the others then the
solution is clearly a black hole microstate. At intermediate
separations such classical distinctions are not meaningful.

\newsec{A Review of the Bubbling Solution}

The eleven-dimensional solutions that have the same
charges and supersymmetries as a three-charge BPS
black hole have the metric
\refs{\GauntlettNW,\GutowskiYV,\BenaDE,\GauntlettQY}:
\eqn\fullmet{\eqalign{ ds_{11}^2& =  - \left({1 \over Z_1 Z_2
Z_3}\right)^{2/3} (dt+k)^2 + \left( Z_1 Z_2 Z_3\right)^{1/3}
h_{mn}dx^m dx^n \cr &+ \left({Z_2 Z_3 \over
Z_1^2}\right)^{1/3}(dx_1^2+dx_2^2) + \left({Z_1 Z_3 \over
Z_2^2}\right)^{1/3}(dx_3^2+dx_4^2) + \left({Z_1 Z_2 \over
Z_3^2}\right)^{1/3}(dx_5^2+dx_6^2) \,,}}
where the four-dimensional base metric $h_{mn}dx^m dx^n$ is
hyper-K\"ahler.  In \refs{\BenaVA,\BerglundVB,\GiustoIP} it was shown
that when $h_{mn}dx^m
dx^n$ is a Gibbons-Hawking (GH) metric:
\eqn\GHmetric{h_{mn}dx^m dx^n ~=~  V^{-1} \, \big( d\psi + \vec{A}
\cdot
d\vec{y}\big)^2  ~+~ V\, (d\vec{y}\cdot d\vec{y} )\,,}
with $\vec y \in \IR^3$ and $ \vec \nabla \times \vec A ~=~ \vec
\nabla V \,$, one can obtain a
smooth eleven-dimensional metric even if the geometric charges
in $V$ are negative. Therefore, we take:
\eqn\Vform{V ~=~   \sum_{j=1}^N \,  {q_j  \over r_j} \,, }
with  $r_j \equiv |\vec{y}-\vec{y}^{(j)}|$.  To be regular at $r_j
=0$ one must choose $q_j  \in \ZZ $, and to be asymptotic to flat
$\IR^4$ one must take ${q_0 ~\equiv~ \sum_{j=1}^N \, q_j  ~=~ 1}$.
The GH metric has non-trivial two-cycles (the bubbles),
$\Delta_{ij}$, defined by the fiber  coordinate, $\psi$, and any
line running between a pair of GH points, $\vec{y}^{(i)}$ and
$\vec{y}^{(j)}$.

The background three-form potentials in the eleven-dimensional solution
are defined via vector potentials in the four-dimensional GH base.
\eqn\Aansatz{
 {\cal A}   ~=~  A^{(1)} \wedge dx_1 \wedge dx_2 ~+~  A^{(2)}   \wedge
dx_3
 \wedge dx_4 ~+~ A^{(3)}  \wedge dx_5 \wedge dx_6\,,}
Supersymmetry requires the  ``dipole field strengths:''
\eqn\Thetadefn{\Theta^{(I)} \equiv d A^{(I)} + d\Big(  {dt +k \over
Z_I}\Big)}
to be self-dual. One can easily show that if the base is
Gibbons-Hawking, the   $\Theta^I$ are given by
\eqn\thetaansatz{\Theta^{(I)} ~=~ - \, \sum_{a=1}^3 \,
\big[\partial_a \big( V^{-1}\, K^I \big)\big] \, \big[ (d \psi
+A)\wedge  dy^a + \coeff{1}{2}\, V \epsilon_{abc} \, dy^b \wedge
dy^c  \big] \,.}
where $K^I$ is harmonic in the $\IR^3$ base of the GH space. To have
 completely non-singular solutions one must choose
these  to be sourced only at the singular points of $V$:
\eqn\KIform{K^I ~=~ \sum_{j=1}^N \, {k_j^I \over r_j} \,.}
One then finds that the two-form fluxes through $\Delta_{ij}$ are given by:
\eqn\fluxperiods{ \Pi^{(I)}_{ij}  ~=~   \bigg( {k_j^I \over q_j} ~-~
{k_i^I \over q_i} \bigg) \,.}

Having chosen the flux parameters, $k_j^I$, the form of smooth
horizonless solutions is completely fixed by the asymptotics and
by requiring the absence of singular sources. See \refs{\BenaVA,
\BerglundVB} for details.  One further important physical
ingredient in the solution is to avoid Dirac-Misner strings and
closed time-like curves (CTC's).  This requires that one to impose the {\it
bubble equations}\foot{One can add a constant to $V$ in \Vform,
and reduce the resulting smooth five-dimensional solution to a singular
four-dimensional multi-black-hole solution of the type explored in
\refs{\DenefNB,\BatesVX}. The  ``bubble equations'' are then equivalent
to the ``integrability conditions'' of \refs{\DenefNB,\BatesVX}. Other
asymptotically four-dimensional configurations
that are smooth in five-dimensions have been explored in
\refs{\BenaAY,\SaxenaUK}} \refs{ \BenaVA, \BerglundVB}:
\eqn\bubbleeqns{ \coeff{1}{6}\, C_{IJK} \, \sum_{{\scriptstyle j=1} \atop
{\scriptstyle j \ne i}}^N \,  \,  \Pi^{(I)}_{ij} \,
\Pi^{(J)}_{ij} \,  \Pi^{(K)}_{ij} \
{q_i \, q_j  \over r_{ij} } ~=~ - \sum_{I=1}^3  \bigg(k^I_i ~-~ q_i \,
\sum_{j=1}^N k_j^I \bigg) \,, }
for $i =1, \dots, N$, and where $r_{ij} \equiv |\vec y^{(i)} - \vec y^{(j)}|$.
The bubble equations are required to remove CTC's in specific,
``potentially dangerous'' limits, and seem to solve the problem globally
in a number of solutions.  However, in general, the absence of CTC's
requires that one ensure that the following are globally true:
\eqn\noCTCs{ V \, Z_I ~ \ge~ 0 \,, \qquad  Z_1\,Z_2 \, Z_3 \,V~-~ \mu^2 \, V^2 ~\ge~ 0 \,,}
where $\mu$ is defined in  \BenaVA.

The charges  of the solution are \refs{\BerglundVB, \newBWW}
\foot{ These results appear to differ by some factors of two
compared to those of \BerglundVB.  This is because our conventions
are those of \BenaVA, which use a different normalization of the
two-form fields.  }:
\eqn\QIchg{Q_I ~=~ -2 \, C_{IJK} \, \sum_{j=1}^N \, q_j^{-1} \,
\tilde  k^J_j \, \tilde  k^K_j\,,}
where
\eqn\ktilde{\tilde  k^I_j ~\equiv~ k^I_j ~-~    q_j\, N  \,  k_0^I  \,
\qquad {\rm and} \qquad k_0^I ~\equiv~{1 \over N} \, \sum_{j=1}^N k_j^I\,.}
The angular momenta are given by:
\eqn\Jright{ J_R ~\equiv~ J_1 + J_2 ~=~ \coeff{4}{3}\, \, C_{IJK} \,
\sum_{j=1}^N q_j^{-2} \, \tilde  k^I_j \, \tilde  k^J_j \,  \tilde  k^K_j \,, \qquad
J_L ~\equiv~ J_1 - J_2 ~=~ 8 \,\big| \vec D\big| \,,}
where
\eqn\dipoles{\vec D_j ~\equiv~  \, \sum_I  \, \tilde k_j^I \, \vec y^{(j)} \,,
\qquad \vec D ~\equiv~ \sum_{j=1}^N \, \vec D_j \,.}

The bubble equations do not seem to allow straightforward and
intuitive solutions.  It is therefore rather natural to see what we
can deduce either without them, or having made some extremely simple
approximations.

\newsec{The Foaming BPS Black Hole}

We first consider a configuration of $N$ GH centers that lie in a
single ``blob'' and that all have roughly the same flux parameters to
leading order in $N$.  To keep the computation simple we take $N=2 M
+1$ to be odd, and consider a distribution with $q_j = (-1)^{j+1}$.
We can take all of $k^I_i$ to be positive numbers, and we will assume
that their variations about their mean value are of the same order as
the mean value:
\eqn\kvariations{k^I_j  ~=~ k_0^I  \, (1 ~+~ \cO(1)) \,,}
where $k_0^I $ is defined in \ktilde.  Using  \QIchg\ we find the
charges to be:
\eqn\charges{{Q_I = 2\, C_{IJK } \Big( N^2 k_0^J k_0^K  -
\sum_j k^J_j k^K_j q_j^{-1}\,   \Big)
\approx~ 2\, C_{IJK }  \big( N^2 + \cO(1)\big )\, k_0^J k_0^K   }}
where we used \kvariations\ and the fact that $|q_i| =1$ only in
the last step.  If $|q_j| \ne 1$ then the last term would yield a
correction of, at most, $ \cO(N)$.

We can make a similar computation for the angular momenta using \Jright:
\eqn\Jfin{\eqalign{J_R &~=~  \coeff{4}{3}\,
C_{IJK} \bigg(2 N^3  k_0^I k_0^J  k_0^K +
\sum_j   q_j^{-2}\,k^I_j k^J_j k^K_j
- 3  N k_0^I  \sum_j  q_j^{-1}\, k^J_j  k^K_j  \bigg) \cr
& ~ \approx~  \coeff{4}{3}\,  C_{IJK}
\big( 2  N^3    + \cO(N ) \big) \, k_0^I k_0^J k_0^K\,,
}}
where we have again used the fact that, for a ``well behaved''
distribution of
positive $k_i^I$ with $|q_j| =1$ and $q_0 =1$  one has:
\eqn\sums{  \sum_i q_i^{-1}  k^J_i  k^K_i ~=~
\sum_i q_i   k^J_i  k^K_i ~\approx~  k_0^J  k_0^K \,, \qquad
\sum_i  k^I_i k^J_i k^K_i ~\approx~ N k_0^I  k_0^J k_0^K \,.}
Once again, if $|q_j| \ne 1$ then  \Jfin\ would remain true to $ \cO(N^2)$.

These results did not involve the solution of the ``bubble
equations.'' However, the formula for $J_L$ does depend rather
non-trivially on the details of such a solution. One can also
estimate the magnitude of $J_L$ in a few increasingly more
complicated situations, and thereby argue that it is typically
subleading in the large-$N$ limit.

Clearly, any configuration with three independent $\ZZ_2$
reflection symmetries will have $\vec D =0$. Hence, if one
distributes all the points on a line (this would give $U(1)\times
U(1)$ invariant microstates),  and if all the $k_i^I$ are equal,
then $\vec D = 0$. If one now takes the points on the line, but
with different $k_i^I$, one can still exactly calculate $\vec D $
\newBWW, and find it to be subleading. Intuitively this is
happening because the largest part of the contributions to $\vec
D$ from the positive charge centers is cancelled by the
contribution from the nearby negative charge centers. The same
intuition can be used to argue that the only way to get a large
$|\vec D|$ from a compact blob is to ``bias it'' in terms of the
locations of, and fluxes on, the positive and negative GH centers.
This is basically what happens in the solutions of \GiustoIP.

If we now focus on M theory on $T^6$, in the large $N$ limit, we have
\eqn\QJ{ Q_1 \approx  4 N ^2 k_0^2 k_0^3 \,, \quad Q_2 \approx
4 N^2 k_0^1 k_0^3 \,, \quad Q_3 \approx 4 N^2 k_0^1 k_0^2 \,; \qquad
 J_R ~\approx~ 16 N^3 k_0^1 k_0^2 k_0^3 \,,}
and hence $J_R^2 ~\approx~ 4 Q_1 Q_2 Q_3 $, exactly as one finds for  the
maximally-spinning BMPV black hole.   One can also estimate how much the
charges of our configurations differ from those of the maximally
spinning BMPV
black hole, and one finds that
\eqn\deviation{{J_R^2 \over 4  Q_1 Q_2 Q_3} ~-~ 1\sim
O\left({1 \over N^2}\right)}

Interestingly enough, the value of $J_R $ is slightly bigger than
$\sqrt{4 Q_1 Q_2 Q_3}$. However, this is not a problem because in
the classical limit, when the black hole solution is valid, these
become equal. One can also argue that a classical black hole with
zero horizon area will receive higher-order curvature corrections,
which typically cause a positive increase in the horizon area;
hence the $S=0$ microstates can have $J_R$ slightly larger than
the maximal classically allowed value in order to compensate for
this increase.

One can also easily see that our formulae reproduce the charges
and angular momentum of the maximally-spinning BPS black hole in
any five-dimensional $U(1)^k$ supergravity. If we define $Y^I$
using
\eqn\ydef{Q_I \equiv {1\over 2} C_{IJK} Y^J Y^K~,}
then the maximal angular momentum of this black hole is
\KalloshVY:
\eqn\othersugra{J_R = {1\over 3} C_{IJK} Y^I Y^J Y^K}
which clearly agrees with \charges\ and \Jfin\ in the large $N$ limit.

\newsec{The Foaming Black Ring}

The next-simplest configuration is a blob of total GH charge zero
with a single, very distant point of GH-charge $+1$.  One can view
this as a continuous deformation of the configuration considered
above.  Again we will assume $N$ to be odd, and take the GH charge
distribution to be $q_j = (-1)^{j+1}$, with the distant charge
being the $N^{\rm th}$ GH charge.

To have the $N^{\rm th}$ GH charge far from the blob means that all
 the two-cycles, $\Delta_{j \, N}$ must support a very large flux
 compared to the fluxes on the $\Delta_{ij}$ for $i,j < N$.  Indeed,
 the former fluxes must be of order $N$ larger than the latter.  To
 implement this, we take:
\eqn\ringblobvars{k^I_j  ~=~ a_0^I  \, (1 ~+~ \cO(1))  \,, \quad j=1,\dots,N -1 \,, \qquad
k^I_N  ~=~ - b_0^I \, N \,.}
where the $b_0^I$ are  parameters of the same order as $a_0^I$, and the
$a_0^I$ are defined as averages of the flux parameters over the blob:
\eqn\ringblobavg{a_0^I ~\equiv~{1 \over N-1}\, \sum_{j=1}^{N-1} \, k_j^I\,.}
The fluxes of this configuration are then:
\eqn\blobfluxes{\Pi^{(I)}_{ij} ~=~ \bigg({ k_j^I \over q_j} ~-~  { k_i^I \over q_i}
\bigg) \,, \qquad \Pi^{(I)}_{i\,N} ~=~  - \Pi^{(I)}_{ N \, i} ~=~ -\bigg(
{ k_i^I  \over q_i} ~+~    N\, b_0^I \bigg) \,, \quad i,j =1, \dots, N-1\,.}
For this configuration one has:
\eqn\abvariations{\eqalign{k_0^I  ~=~ & {(N-1)\over N} \, a_0^I ~-~b_0^I \,,
\qquad \tilde k_N^I  ~=~  - (N-1) \, a_0^I   \,, \cr
\qquad  \tilde k^I_j  ~=~ & a_j^I ~+~ q_j \, (N\, b_0^I  - (N-1)\, a_0^I)\,, \quad
j=1,\dots, N-1 \,.}}
Motivated by the bubbling black ring of \BenaVA, define the physical parameters:
\eqn\dandf{d^I ~\equiv~  2\,(N-1)\, a_0^I \,, \qquad f^I ~\equiv~
(N-1)\, a_0^I - 2\,N\, b_0^I \,.}
Then exactly the same arguments as those employed earlier lead to
the following expressions for the charges and angular momentum:
\eqn\ringchgs{\eqalign{Q_I ~=~ &(1 + \cO(N^{-2}) )\, C_{IJK} d^J f^K   \,, \cr
J_R  ~=~ & (1 + \cO(N^{-2}))\, \Big(\coeff{1}{2} \, C_{IJK}
(d^I d^J f^K + f^I f^J d^K) ~-~ \coeff{1}{24} \, C_{IJK}  d^I d^J d^K \Big) \,.}}

Since the $N^{\rm th}$ point is far from the blob,
we can take $r_{i N} \approx r_0$ and then the $N^{\rm th}$ bubble
equation reduces to:
\eqn\bubblerad{\coeff{1}{6}\, C_{IJK} \,\sum_{j=1}^{N-1}\,
\bigg({a_j^I \over q_j} + N\, b_0^I\bigg)\, \bigg({a_j^J \over q_j} + N\, b_0^J\bigg)\,
\bigg({a_j^K \over q_j} + N\, b_0^K\bigg)\, {q_j  \over r_0}~=~  (N-1)\, \sum_I \, a^I\,.}
To leading order in $N$ this means that the distance from the
blob to the $N^{\rm th}$ point, $r_0$, in the GH space is given by:
\eqn\ringrad{r_0 ~\approx~   \coeff{1}{2} \, N^2\, \bigg[\sum_I \, a^I\bigg]^{-1}
C_{IJK} \,   a_0^I \, b_0^J \, b_0^K ~=~  \coeff{1}{32} \,
\bigg[\sum_I \, d^I\bigg]^{-1}  C_{IJK} \,   d^I \, (2f^J - d^J) \, (2f^K - d^K) \,.}

Finally, considering  the dipoles, \dipoles, it is evident that to leading
order in $N$, $\vec D$ is dominated by the contribution coming from the
$N^{\rm th}$ point and that:
\eqn\ringJL{\eqalign{J_1 - J_2 ~=~ & 8\, | \vec D| ~\approx~  8\, (N-1) \, \bigg(
\sum_I \, a_0^I \bigg)\,  r_0 ~\approx~ 4 \, N^2\, (N-1) \, C_{IJK} \,   a_0^I \,
b_0^J \, b_0^K   \cr  ~=~ & \coeff{1}{8} \,
C_{IJK} \,   d^I \, (2f^J - d^J) \, (2f^K - d^K) \,.}}

One can easily verify that these results perfectly match the
properties of the bubbled supertube found in \BenaVA. Thus the
blob considered here  has  exactly the  same size, angular momenta
and charges as the zero-entropy three-charge black ring.

\newsec{The Entropy of the Foam}

As we have seen, when the number of centers is very large, getting
a solution with $J_1^2=J_2^2= Q_1 Q_2 Q_3$ is not hard. In fact,
it seems to be much harder to  get anything else, unless one
drastically increases the flux parameters $k_i^I$ on some subset
of the cycles. It is natural therefore to ask how many bubbled
black hole configurations are there for some given total charges.

If we work in M theory on $T^6$, then \QJ\ implies that
\eqn\sumk{\sum k^1_i = {1 \over 2}  \sqrt{Q_2 Q_3 \over Q_1} ~,}
and similarly for $ \sum k^2_i$ and $\sum k^3_i $.  Note that these
relations depend upon neither the number nor the charges of the GH
points.  Hence, there is a non-trivial entropy coming simply from the
many possibilities of choosing the positive $k_i^I$ subject to
constraints of the form \sumk.  Since this entropy comes entirely from
the combinatorics of laying out quantized fluxes on topologically
non-trivial cycles, we will refer to it as the {\it topological
entropy}.

\subsec{Topological Entropy}

The parameters $k^I_i$ are half-integers (up to a gauge
shift induced by $K^I \to K^I + c^I V$). This can be seen most
easily from their relation to the integer black ring dipole charge
\dandf, and can be also derived by requiring the integral of
$\Theta^{(I)}$ over the $S^2$ surrounding any GH center in $\IR^3$
to be quantized.   The entropy in choosing the positive half-integers
$k^1_i$ is therefore:
\eqn\entropy{S =2 \pi \sqrt{{1\over 6} {\left(Q_2\, Q_3 \over
Q_1\right)} ^{1/2} }\,, }

Naively there should be similar factors coming from partitioning $k^2_i$ and $k^3_i$,
which leads to:
\eqn\stotal{S_{topological} = 2  \pi \left( \sqrt{{1\over 6}
{\left(Q_2\, Q_3 \over Q_1\right)} ^{1/2} } +
 \sqrt{{1\over 6} {\left(Q_1 Q_2 \over Q_3\right)} ^{1/2} }
+
 \sqrt{{1\over 6} {\left(Q_1 Q_3 \over Q_2\right)} ^{1/2} }~
\right)}
There are, however some subtleties.  First, the partitioning of
$k^1_i$, $k^2_i$ and $k^3_i$ is not completely independent.  A bubble
will collapse unless all three fluxes are non-zero, and so we should
count the ways of having non-zero partitions of all the $k^I_i$ over
$N$ bubbles and then sum over $N$.  One can show that this constraint
only contributes only sub-leading ($\log(Q)$) corrections to the
entropy (see \newBWW\ for details).  Secondly, given the $k^1_j$,
there are also further constraints on $k^2_j$ and $k^3_j$ imposed by
the global absence of CTC's.  These conditions are somewhat more
difficult to handle but we believe that the bubble equations, combined
with some suitable positivity conditions on the fluxes, will suffice
to guarantee  the conditions in \noCTCs, and hence that \stotal\ is
correct to  leading order.
Independent of these subtleties and the constraints on $k^2_j$ and
$k^3_j$ for a given set of $k^1_j$, we see from \entropy\ alone that
the topological entropy grows as $Q^{1/4}$.

If the topological entropy dominates then the ``typical''
microstate has of order $Q^{1/4}$ centers, with flux parameters of
order $Q^{1/4}$ on each.  One can estimate the typical distance
(measured in the full, physical eleven-dimensional metric) between
two adjacent points and one finds that it is of order $Q^{1/4}$,
which is parametrically larger than the Planck scale.

There exists another possible source of topological entropy.  We have
shown that, for a large number of bubbles, the charges \charges\ and
angular momenta \Jfin\ only depend on the $k_i^I$, and not on the
$q_i$, and so it appears that any ``well-behaved'' distribution of
$q_i$ will give a good microstate. Since the $q_i$ can be arbitrary
positive and negative integers, the number of such configurations is
infinite.  However, one must recall that the quantization of fluxes on
the two-cycles running between any two GH centers implies that the
$\Pi_{ij}^{(I)}$ are half-integers\foot{This quantization condition is
somewhat more stringent than that suggested in
\BerglundVB. One can also try to estimate the topological entropy using the quantization in \BerglundVB. At first glace it appears 
to only differ from \entropy\ by terms that are subleading in the charges.}. Therefore, for a certain distribution of the
$k_i^I$, the $q_i$ can only take a finite set of values and so the
contribution to the topological entropy coming from having $|q_i| \ne
1$ is subleading. Alternatively, this quantization rule means that, up
to a gauge shift, the $k_j^I/q_j$ are half-integers, and so there are
no ``frozen'' GH points: It is possible to separate GH charges at a
point with $|q_j| \ne 1$ and partition the fluxes while respecting the
quantization rules. Thus the configurations with $|q_j| \ne 1$ are
merely special points in the phase space of solutions with $|q_j| =
1$.

\subsec{Moduli Space Entropy}

Another potentially large source of entropy
arises from the moduli associated with the possible locations of the
GH centers: There are $3(N-1)$ such positions (excluding the center of
mass) and $(N-1)$ constraints from the bubble equations, \bubbleeqns.
In \BenaVA\ this moduli space was analyzed in simple examples and it
was found that, in the first approximation, each cycle running between
adjacent positive and negative centers behaves like an almost rigid
rod.  The most generic configuration is one in which all the rods have
different orientations.  It is not clear how to quantize this moduli
space and thereby compute the entropy, but one can give a very rough estimate.

The length, $L$, in $\IR^3$ of a rod, or bubble, (measured
in the full, physical eleven-dimensional metric) is of order the flux,
$k$, through the bubble.  The phase space of such a rod should be
proportional to the number of Planck size regions on a two-sphere of
radius $L$, which is approximately $L^2/l_p^2 \sim k^2$.  There are $N
\sim \sqrt{Q}/k$ rods in such a configuration, and so the number of
states of charge $Q$ and flux $k$ is approximately:
\eqn\roddegen{ d(Q,k) ~\sim~ k^{2\,{ \sqrt{Q} \over k}} \,.}
As a function of $k$, this has a  maximum at $k = e$
(in Planck units) and this gives a contribution to the entropy of
\eqn\Smod{ S_{moduli} ~\sim~ {2\over e} \, \sqrt{Q}\,.}

Our rough estimate indicates therefore that this contribution to the
entropy is dominated by Planck scale bubbles, and is much larger than
the topological entropy.  It remains to be seen whether this rather
naive analysis of the the quantum moduli space is valid and whether
the description of the bubbles in terms of ``rods'' is sufficiently
accurate to determine the dominant contribution to the entropy.  A
more involved analysis of the moduli space entropy for four-dimensional
multi-black-hole solutions has been performed in \DenefRU. It would be
interesting to see if one could adapt and extend that analysis to the
present problem.

While both the moduli space entropy and the topological entropy
are large, they are not large enough to be classically visible in
supergravity; this is not surprising, since the classical entropy
of the black hole with the same charges is zero\foot{Nevertheless,
it is quite possible that there might exist a ``small black hole''
whose entropy matches \stotal\ or \Smod.}.

Finally, we note that the geometry of the distribution of points
in the $\IR^3$ of the GH base can be deceptive when it comes to
understanding the full geometry of the microstates. Even if  all
the points are very close together  (in the metric in $\IR^3$) and
appear to be sitting on a long line or in a blob, the warp factors
serve to blow up the bubbles and expand the space-time metric to
create the familiar BPS black hole throat. Indeed, if one draws an
$S^2$ around all the GH points (and thus an $S^3$ around the
bubbles), then outside this sphere the geometry will look very
like the throat of a rotating BPS  black hole.  This is because
the formation of the throat is a consequence of the net charges
within the circumscribing sphere and these are the same for  the
bubbled microstates and for the black hole. The only difference is
inside this sphere:  The black hole has an infinite throat, while
the microstate geometries have a finite one.

In a forthcoming paper \newBWW\ we will examine the appearance of
the bubbled geometries in more detail and perform several gedanken
experiments involving merging black ring foams and black hole
foams and comparing with the results of \BenaZY.  This will
further establish the fact that the foam has the same size as the black
hole.

\newsec{Conclusions and Future Directions}

We have explored the charges and angular momenta of a large family of
smooth three-charge BPS geometries. We have found that, in the limit
when the number of two-cycles is large, a large subclass of these
geometries have the correct charges, angular momenta, and size to be
microstates of the three-charge maximally-spinning BPS black hole in
five dimensions.

A remarkable result of this analysis is that the upper bounds on
the BPS black hole angular momenta, $J_1=J_2= \sqrt{Q_1 Q_2 Q_3}$,
which are classically found by requiring the absence of closed
time-like curves on the black hole horizon, also emerge from the
study of smooth horizonless solutions that represent black hole
microstates.  This is also not a special feature of M-theory on $T^6$;
it also true in any $U(1)^k$ supergravity.

In fact, using a GH base with a large number of alternating plus
and minus centers localized in a single ``blob,'' and with a
well-behaved distribution of flux parameters, it is hard  to
obtain anything but maximally-spinning BPS black hole microstates.
It would be interesting to try to see if one can obtain microstate
geometries with $J_1=J_2 < \sqrt{Q_1 Q_2 Q_3}$ using a GH base. It
is possible that such solutions might only be obtained using a
less-constrained ({\it i.e.} not Gibbons-Hawking)
four-dimensional hyper-K\"ahler  manifold as a
base.

All the solutions obtained here can be easily be dualized to a
frame where they have D1, D5, and P charges, and can be made
asymptotic to $AdS^3 \times S^3 \times T^4$ \BenaTK.  Hence, they
are dual to states of the D1-D5 CFT. It is very important to try
to understand what these CFT states are; this might allow us to
find whether the typical black-hole CFT microstates are dual to
bubbling solutions.

Another interesting result of our investigation is that the
microstates of a  zero-entropy black ring and the  microstates of
a zero-entropy black hole are all part of one and the same family:
One can simply interpolate between them by moving a GH center with
charge $+1$  away from the other GH centers, and there is no clear
boundary between them. The fact that black-ring and black-hole
microstates belong to the same family of solutions is very
satisfying from the perspective of the micro-physics of black
objects.  It also indicates that the dual states in the D1-D5 CFT
are very similar, and this might represent an important step
towards finding what these states are.

A third result of our investigation is that there are two
significant contributions to the entropy of solutions: The
topological entropy and the entropy from the moduli space. Our
rough estimates suggest that the latter will dominate and that a
typical geometry will have the size of the black hole, but will
contain Planck-size bubbles. On the other hand, it is possible
that our estimates were rather too simplistic.  If the topological
entropy were dominant, then the typical geometry would have
bubbles and curvatures parametrically lower than the Planck scale.
It is therefore very important to understand how to address the
quantization of this moduli space.

It should also be remembered that our analysis here is based upon
geometries with a Gibbons-Hawking base metric, which are a very
special class of the allowed hyper-K\"ahler base metrics.
This was done primarily because they enable us to perform
explicit computations.  From the physics of supertubes
\refs{\BenaWT,\BenaDE,\BenaTD} we expect more general
hyper-K\"ahler metrics to involve a large number of arbitrary
functions \BenaVA, which should give a contribution to the entropy
much larger than the topological and the moduli-space entropies. It is
very important to build and understand these metrics, and to find
whether this hyper-K\"ahler entropy favors large or small bubbles.

Finally, this work provides a starting point from which one can
perform an entropy calculation of the black hole microstates in
the spirit of \refs{\PalmerGU,\Bak}, using perhaps the techniques of
\refs{\GrantQC,\RychkovJI} which could prove or disprove Mathur's
conjecture. Such calculations are usually done by perturbing
around classical zero-entropy configuration and, prior to the our
bubbled solutions, no such black hole microstates were known. This
obstacle to progress has now been eliminated.

\bigskip
\leftline{\bf Acknowledgments}

We would like to thank Juan Maldacena, Don Marolf, Per Kraus, Slava
Rychkov and Davide Gaiotto for interesting discussions. I.B. would
like to thank the Kavli Institute for Theoretical Physics for
hospitality during part of this work. The work of NW and CWW is
supported in part by the DOE grant DE-FG03-84ER-40168.  The work of IB
is supported in part by the NSF grants PHY-0503584 and PHY-990794.

\listrefs
\end